\documentclass[aps,prl,twocolumn,superscriptaddress,showpacs,preprintnumbers,amsmath,amssymb]{revtex4}
\usepackage{graphicx,amssymb}

\begin{document}

%

\preprint{Ver 0.4}

\title{Direct observation of mixing of spin-multiplets in an antiferromagnetic molecular nanomagnet by electron paramagnetic resonance}

\author{S. Datta}
\affiliation{Department of Physics, University of Florida, Gainesville, Florida 32611, USA}

\author{O. Waldmann}
\affiliation{Department of Chemistry and Biochemistry, University of Bern, CH-3012 Bern, Switzerland}

\author{A. D. Kent}
\affiliation{Department of Physics, New York University, 4 Washington Place, New York, New York 10003, USA}

\author{V. A. Milway}
\author{L. K. Thompson}
\affiliation{Department of Chemistry, Memorial University, St. John's, Newfoundland, Canada A1B 3X7}

\author{S. Hill}
\affiliation{Department of Physics, University of Florida, Gainesville, Florida 32611, USA}

\date{\today}

\begin{abstract}
High-frequency electron paramagnetic resonance (EPR) studies of the
antiferromagnetic Mn-$[3\times 3]$ molecular grid clearly reveal a
breaking of the $\Delta S = 0$ selection rule, providing direct
evidence for the mixing of spin wavefunctions ($S$-mixing) induced
by the comparable exchange and magneto-anisotropy energy scales
within the grid. This finding highlights the potential utility of
EPR for studies of exchange splittings in molecular nanomagnets,
which is normally considered the sole domain of inelastic neutron
scattering, thereby offering improved sensitivity and energy
resolution.
\end{abstract}

\pacs{33.15.Kr, 71.70.-d, 75.10.Jm}

\maketitle

%

The synthesis of ordered crystalline arrays (ensembles) of nominally identical magnetic molecules has enabled
unprecedented insights into quantum magnetization dynamics at the nanoscale, leading to exciting discoveries
such as the quantum interference associated with the coherent rotation (tunneling) of a large spin
\cite{Mn12_Fe8_etc}. The molecular approach is particularly attractive due to the highly ordered and
monodisperse nature of the molecules in the solid state, and because one can systematically vary many key
parameters that influence the quantum behavior of a magnetic molecule, e.g., the total spin, symmetry, etc.
The magnetism of these molecules can be described quite generally by the microscopic spin Hamiltonian
\cite{dipdip}
\begin{equation}
\hat{H} = -\sum_{ij}{ J_{ij}\hat{\textbf{s}}_i \cdot \hat{\textbf{s}}_j } +
\sum_{i}{\hat{\textbf{s}}_i \cdot \textbf{D}_i \cdot \hat{\textbf{s}}_i } + g \mu_B \hat{\textbf{S}} \cdot
\textbf{B}.
\end{equation}
The first term parameterizes the Heisenberg interaction between each pair of magnetic ions;
$\hat{\textbf{s}}_i$ are the spin operators at each site, and $J_{ij}$ the respective coupling constants. The
second term accounts for the magnetic anisotropy at each site; $\textbf{D}_i$ are the local
zero-field-splitting (ZFS) tensors. The final term is the Zeeman interaction written in terms of the total
spin operator $\hat{\textbf{S}}$.

In the strong exchange limit, the total spin $S$ is a good quantum number and the exchange term splits the
energy spectrum into well separated spin multiplets, each having comparatively weaker ZFS due to magnetic
anisotropy \cite{Bencini90}. However, as the ZFS increases relative to the exchange splitting, mixing of spin
multiplets becomes significant ($S$-mixing) such that $S$ no longer constitutes a good quantum number
\cite{Liviotti02,OW_HighSpin}. This parameter regime has attracted growing interest due to the realization that
$S$-mixing plays a crucial role in the quantum dynamics of coupled spin systems, namely in the tunneling
terms in single-molecule magnets (SMMs) and antiferromagnetic (AF) clusters
\cite{Prokofev98,Carretta04,SH_Ni4,OW_NVT}, and in the novel magneto-oscillations associated with the total
spin of a molecule \cite{OW_Mn3x3QMO,Carretta03,Carretta07}. Indeed, the interplay between anisotropy and
exchange is central to the understanding of (nano-) magnetism, and experimental determination of exchange and
anisotropy splittings is of great value.

ZFS caused by magnetic anisotropy may be studied spectroscopically
using both electron paramagnetic resonance (EPR) and inelastic
neutron scattering (INS). In contrast, the most unambiguous method
for estimating exchange couplings involves determining the exact
locations of excited spin multiplets. For this reason, such
investigations have been limited to INS, since the EPR selection
rule $\Delta S = 0$ forbids excitations between spin multiplets (the
INS selection rule $\Delta S = 0,\pm1$ permit such
\emph{inter}multiplet transitions). In this work, we show that
$S$-mixing indeed can give rise to a situation in which exchange
splitting is observed directly in a nanomagnet using
high-sensitivity multi-high-frequency EPR.

\begin{figure}[t]
\includegraphics[scale=0.88]{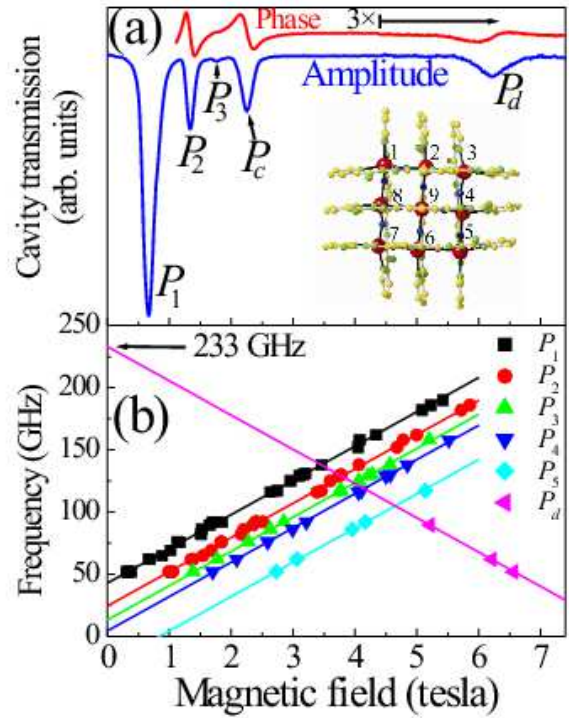}
\caption{\label{fig:1} (Color online) (a) Complex EPR spectrum at 1.4~K and 62~GHz. The dips in amplitude
correspond to EPR; the sense of phase rotation through each dip differentiates $\Delta M = \pm 1$
transitions. The Mn-$[3\times 3]$ grid is shown in the inset. (b) Frequency dependence of the resonance
fields $B_n$ of the indicated peaks; the solid lines represent best-fits (see text).}
\end{figure}

The Mn-$[3\times 3]$ grid is an attractive candidate in the above context. Its structure consists of nine
spin-$\frac{5}{2}$ Mn$^{\rm II}$ ions placed at the vertices of a $3\times3$ matrix, see Fig.~\ref{fig:1}(a).
A significant magnetic anisotropy, demonstrated in previous experiments \cite{Guidi04}, gives rise to several
striking effects, including novel quantum-oscillations in the field dependent magnetic torque
\cite{OW_Mn3x3QMO}, and tunneling of the N\'eel vector at high fields \cite{OW_Mn3x3NVT}. The magnetism of
Mn-$[3\times 3]$ can be well described by the approximate Hamiltonian
\begin{eqnarray}
\hat{H}_1 &=& - J_R \left( \sum^7_{i=1}{ \hat{\textbf{s}}_i \cdot \hat{\textbf{s}}_{i+1} } + \hat{\textbf{s}}_8
\cdot \hat{\textbf{s}}_1 \right)
- J_C \sum^4_{j=1}{ \hat{\textbf{s}}_{2j} \cdot \hat{\textbf{s}}_9}
\cr
&&+ D_R \sum^8_{i=1} \hat{s}^2_{i,z} + D_C \hat{s}^2_{9,z}
+ g \mu_B \hat{\textbf{S}} \cdot \textbf{B},
\end{eqnarray}
with AF couplings $J \equiv J_R = J_C<0$, and uniaxial easy-axis
anisotropy $D \equiv D_R = D_C <0$ (spins are numbered according to
Fig.~\ref{fig:1}; $z$ denotes the axis perpendicular to the grid)
\cite{OW_Mn3x3QMO}. Here, we report the clear observation of an EPR
transition between the $S=\frac{5}{2}$ ground state and the first
excited $S=\frac{7}{2}$ spin multiplet, i.e., a $\Delta S = \pm 1$
transition.

The observation via EPR of \emph{inter}multiplet transitions for a molecular nanomagnet is of significance
from at least three points of view.

(1) It demonstrates that with today's EPR spectrometers it is possible to directly measure exchange
splittings. This represents a significant development, with the potential to change the landscape of
experiments on exchange coupled systems. EPR offers many advantages compared to INS, such as high-sensitivity
(much smaller single crystals), "cheap" experimental environments, essentially unlimited spectral resolution
permitting linewidth studies, etc..

(2) To date, the most direct evidence for $S$-mixing has involved measuring energy spacings in EPR and INS
spectra, and then comparing these to calculations \cite{SH_Ni4,Carretta07}. In the present case, however,
evidence comes from the observation of an otherwise forbidden EPR transition, i.e., breaking of the selection
rule $\Delta S = 0$ $-$ a property directly related to the wavefunctions. This is, hence, the most clear-cut
experimental demonstration of $S$-mixing.

(3) Mn-$[3\times 3]$ was the first molecule in which the effects of $S$-mixing were clearly demonstrated, and
subsequent studies have shown that the generic Hamiltonian $\hat{H}_1$ provides a complete description of
these effects \cite{OW_Mn3x3QMO}. The present EPR results confirm these findings and provide text-book
quality insights into the physics of magnetic molecules and the effects of $S$-mixing.

%

Single crystals of Mn-$[3\times 3]$, [Mn$_9$(2POAP-2H)$_6$] (ClO$_4$)$_6$$\cdot$3.57MeCN$\cdot$H$_2$O, were
prepared as reported \cite{Zhao00}. They crystallize in the space group $C_2/c$. The cation
[Mn$_9$(2POAP-2H)$_6$]$^{6+}$ exhibits a slightly distorted $S_4$ molecular symmetry, with the $C_2$ axis
perpendicular to the grid plane and parallel to the uniaxial magnetic easy ($z$-) axis. Single-crystal EPR
spectra were obtained using a sensitive cavity perturbation technique. A vector network analyzer was used to
record the complex signal (amplitude and phase) transmitted through the cavity; this technique is described
in detail elsewhere \cite{HFEPR}. Data were recorded for two samples. If not otherwise stated, the magnetic
field was applied along the $z$ axis.

%

\begin{figure}[t]
\includegraphics[scale=0.7]{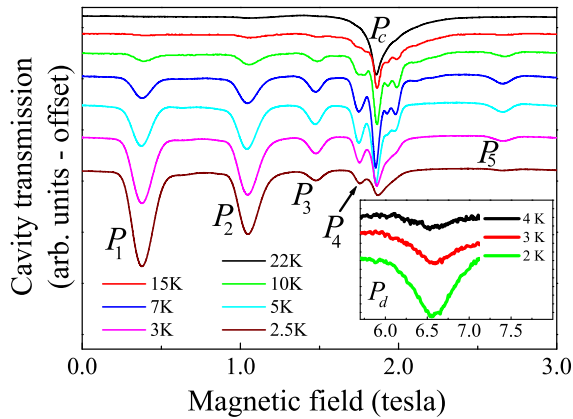}
\caption{\label{fig:2} (Color online) Temperature dependence of the EPR spectrum at 52~GHz; several of the
transitions have been labeled. The inset shows the temperature dependence of transition $P_d$.}
\end{figure}

Figure~\ref{fig:1}(a) displays the 62~GHz EPR spectrum obtained at 1.4~K. Five transitions labeled $P_n$ ($n
= 1, 2, 3, c, d$) are clearly observed. $P_c$ is the $M = -\frac{1}{2} \rightarrow + \frac{1}{2}$ signal
expected for a half-integer spin system, and will not be discussed in detail. The temperature dependence of
the 52~GHz EPR spectrum is shown in Fig.~\ref{fig:2}. Upon increasing the temperature, additional transitions
appear, e.g., $P_4$, $P_5$, and the rich fine structure around $P_c$, which is of no interest in this work.
Above $\sim 20$~K, only a central line is observed, which is expected due to the thermal population of a
large number of spin multiplets with negligible ZFS. In Fig.~\ref{fig:1}(b), we display the frequency
dependence of the resonance fields $B_n$ of the peaks $P_n$ ($n=1, \ldots, 5, d$); each depends linearly on
frequency due to the Zeeman term in Eq.~(2). The ZFS may be obtained from the $B=0$ intercepts by
extrapolation. $B_1$ to $B_5$ increase with frequency; in contrast, $B_d$ decreases with frequency as also
seen from the sense of the phase rotation through the resonance [Fig.~\ref{fig:1}(a)]. A clear hint that
$P_d$ is not a usual intramultiplet transition ($\Delta S = 0$) is found from the large $B=0$ offset of $\sim
233$~GHz. ZFS caused by magnetic anisotropy is much smaller for Mn$^{\rm II}$ complexes \cite{Boca}. This
suggests that the large offset for $P_d$ is caused by exchange, i.e., $P_d$ is an \emph{inter}multiplet
transition. Furthermore, previous magnetic and INS measurements determined the zero-field gap between the $S
= \frac{5}{2}$ ground and first excited $S = \frac{7}{2}$ states to be $230\pm10$~GHz
\cite{OW_Mn3x3QMO,Guidi04}, thus suggesting that $P_d$ indeed corresponds to the $S = \frac{5}{2} \rightarrow
\frac{7}{2}$ transition.

%

Before discussing $P_d$ further, we focus on the analysis in terms of $\hat{H}_1$, with the aim to obtain
best-fit values for $J$ and $D$. In a first step, we fit straight lines to the data in Fig.~\ref{fig:1}(b)
with a common $g$ factor, yielding values for the ZFS associated with each transition (see Table~\ref{tab:1})
and $g$ = 1.970(7). We then determine $J$ and $D$ as follows. A least-squares fit based on the full
Hamiltonian would be very demanding because its dimension is 10,077,696. However, it has previously been
demonstrated that the energies and wavefunctions of the low-lying states of relevance here can be calculated
with high accuracy from the effective spin Hamiltonian $\hat{H}_2 = - 0.526 J_R \hat{\textbf{S}}_A \cdot
\hat{\textbf{S}}_B - J_C \hat{\textbf{S}}_A \cdot \hat{\textbf{s}}_9 + 0.197 D_R ( \hat{S}^2_{A,z} +
\hat{S}^2_{B,z} ) + D_C \hat{s}^2_{9,z} + g \mu_B \hat{\textbf{S}} \cdot \textbf{B}$ \cite{OW_Mn3x3NVT}. $A$
($B$) denotes the sublattice of corner (edge) spins [see Fig.~\ref{fig:1}(a)], i.e., $\hat{\textbf{S}}_A =
\hat{\textbf{s}}_1+\hat{\textbf{s}}_3+\hat{\textbf{s}}_5+\hat{\textbf{s}}_7$ ($\hat{\textbf{S}}_B =
\hat{\textbf{s}}_2+\hat{\textbf{s}}_4+\hat{\textbf{s}}_6+\hat{\textbf{s}}_8$). The dimension of $\hat{H}_2$
is only 2646, permitting a true least-squares fit to the data. We obtain $J = - 4.76(4)$~K and $D =
-0.144(2)$~K, in very good agrement with previous experiments \cite{OW_Mn3x3QMO,Guidi04}. The calculated ZFS
values are compared to experiment in Table~\ref{tab:1}; the agreement is excellent.

\begin{table}
\caption{\label{tab:1} First line: Experimental ZFS determined from
fits to the data in Fig.~\ref{fig:1}(b). Second line: Calculated ZFS
using the best-fit $J$ and $D$ values. Energies are given in GHz.}
\begin{ruledtabular}
\begin{tabular}{lcccccc}
$P_1$ & $P_2$ & $P_3$ & $P_4$ & $P_5$ & $P_d$ \\
\hline
42.7(3) & 24.3(4) & 13.2(5) & 4.4(5) & -23.3(7) & 233(1) \\
42.80 & 24.47 & 12.62 & 3.79 & -24.47 & 233.0
\end{tabular}
\end{ruledtabular}
\end{table}

From a general point-of-view, our model for Mn-$[3\times 3]$ appears
over-simplified. Even if one assumes ideal symmetry, the exchange
and anisotropy parameters need not be identical for all Mn$^{\rm
II}$ ions in the Mn-$[3\times 3]$ grid. Indeed, evidence for slight
variations in the exchange constants was inferred from INS studies
\cite{Guidi04}. However, no variation was found from the present
measurements, or from thermodynamic studies \cite{OW_Mn3x3QMO}. This
is because the relevant energies are governed by the average of the
exchange constants, hence only a single $J$ is needed. Similar
arguments apply to the anisotropy parameters and, in fact, no
evidence for a variation has been reported so far. Consequently, the
generic Hamiltonian $\hat{H}_1$ (or $\hat{H}_2$) with just the three
parameters $J$, $D$, and $g$, captures all of the relevant physics
and provides an excellent effective description of the low-energy
properties of Mn-$[3\times 3]$. The EPR results presented here
establish the most critical test so far.

\begin{figure}
\includegraphics[scale=0.9]{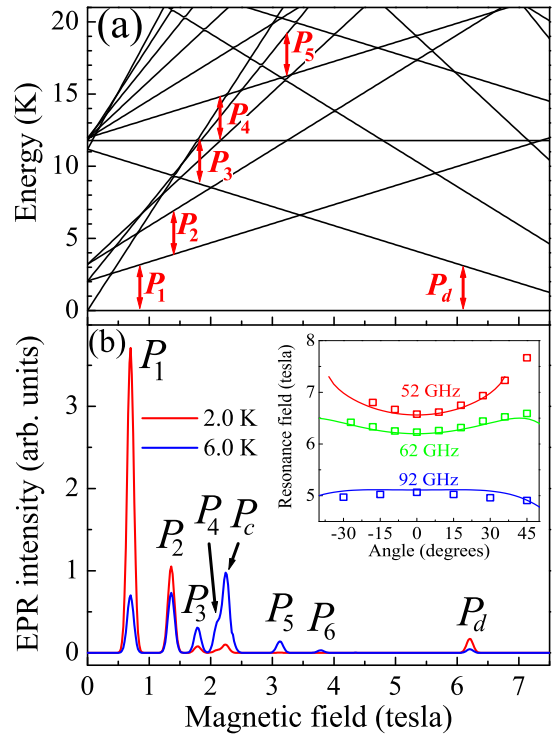}
\caption{\label{fig:3} (Color online) (a) Calculated energy spectrum (relative to the $S = \frac{5}{2}$, $M =
-\frac{5}{2}$ ground state) for $B \parallel z$; several transitions are indicated. (b) Calculated EPR
spectra at 62~GHz and two temperatures. The inset shows the angle dependence of $B_d$ at three frequencies;
experimental data are represented by open squares, and simulations by solid lines.}
\end{figure}

The calculated low-energy, low-field spectrum for Mn-$[3\times 3]$
is displayed in Fig.~\ref{fig:3}(a). It consists of a
$S=\frac{5}{2}$ ground state and a first excited $S=\frac{7}{2}$
multiplet; these are further split into three and four $\pm M$
sublevels, respectively. In zero field, the $M=\pm \frac{5}{2}$
($\pm \frac{7}{2}$) levels of the $S=\frac{5}{2}$ ($S=\frac{7}{2}$)
multiplet lie lowest in energy due to the easy-axis anisotropy. The
$\pm M$ sublevels split in magnetic field due to the Zeeman
interaction. The $M = -\frac{5}{2} \rightarrow -\frac{3}{2}$, $M =
-\frac{3}{2} \rightarrow -\frac{1}{2}$, and $M = \frac{1}{2}
\rightarrow \frac{3}{2}$ transitions within the $S = \frac{5}{2}$
multiplet give rise to $P_1$, $P_2$, and $P_5$, respectively. Hence,
the ZFS for $P_2$ and $P_5$ should be equivalent in magnitude, but
opposite in sign, in agreement with observation (see
Table~\ref{tab:1}). Peaks $P_3$ and $P_4$ correspond to the $M =
-\frac{7}{2} \rightarrow -\frac{5}{2}$ and $M = -\frac{5}{2}
\rightarrow -\frac{3}{2}$ transitions within the $S = \frac{7}{2}$
multiplet. As already noted, $P_d$ corresponds to the transition
between the $M=-\frac{5}{2}$ level of the $S=\frac{5}{2}$ multiplet
and the $M=-\frac{7}{2}$ level of the $S=\frac{7}{2}$ multiplet. The
observed temperature dependence (Fig.~\ref{fig:2}) is perfectly
consistent with these assignments.

In the inset to Fig.~\ref{fig:3}(b), we display the field-orientation-dependence of $B_d$ at several
frequencies. Agreement between experiment (open squares) and theory (solid curves) is very good, including
the opposing trends at lower and higher frequencies/fields. This behavior can be traced to the level
repulsion ($S$-mixing) between the $S=\frac{7}{2}$, $M=-\frac{7}{2}$ level and various $S=\frac{5}{2}$ states
as the field is tilted away from the symmetry direction of the Mn-$[3\times 3]$ grid. As such, this
non-linear frequency dependence of $B_d$ represents evidence for quantum spin-state oscillations
\cite{Carretta07}.

According to the above, $P_d$ violates the $\Delta S = 0$ selection rule for EPR. However, labeling levels by
$S$ can be misleading because of strong $S$-mixing. In fact, $S$ is no longer a good quantum number, though
we retain the notation as a matter of convenience. In order to rigorously check the peak assignments, we
performed a full simulation of the EPR spectrum. Results for 62~GHz are shown in Fig.~\ref{fig:3}(b) at two
temperatures. The agreement with experiment is once again excellent. Most importantly, the calculations
confirm the significant intensity of $P_d$. Hence, the mixing of spin multiplets is strong enough in
Mn-$[3\times 3]$ to observe $P_d$ using state-of-the-art high-frequency EPR techniques ($P_6$ is not observed
experimentally due to its 10 times lower calculated intensity). The wavefunctions of the two states involved
in $P_d$ are calculated as $0.9861 |\frac{5}{2},-\frac{5}{2}\rangle + 0.1625 |\frac{7}{2},-\frac{5}{2}\rangle
- 0.0343 |\frac{9}{2},-\frac{5}{2}\rangle + \ldots$, and $0.9922 |\frac{7}{2},-\frac{7}{2}\rangle -0.1213
|\frac{9}{2},-\frac{7}{2}\rangle + 0.0270 |\frac{11}{2},-\frac{7}{2}\rangle + \ldots$ (with an $|S,M\rangle$
notation for the basis functions). Hence, the ground state has 16\% of $S = \frac{7}{2}$ admixed to it, or
2.6\% in squared (intensity) units.

In the strong exchange limit ($|J|>>|D|$), the zero-field energies within each multiplet are expected to
scale with $M$ as $M^2$. However, this behavior is strongly perturbed in Mn-$[3\times 3]$, with the $M= \pm
\frac{3}{2}$ and $\pm \frac{1}{2}$ states even occurring out of sequence for the $S=\frac{7}{2}$ multiplet! A
standard approach involves the use of an effective spin Hamiltonian, $\hat{H}_{S} = D \hat{S}_z^2 + B^0_4
\hat{O}^0_4$, to describe each spin multiplet. For a $S=\frac{5}{2}$ state, deviations from the expected
$M^2$ behavior may be captured by the $B^0_4$ parameter (for larger $S$, higher order terms such as $B^0_6
\hat{O}^0_6$ may be important also). It is the $S$-mixing that gives rise to significant $B^0_4$ values (and
higher order terms), as noted in previous works \cite{Liviotti02,OW_HighSpin,Pilawa03,SH_Ni4}. The ratio
$|B^0_4/D|$, therefore, serves as a measure of the degree of $S$-mixing. For the $S=\frac{5}{2}$ multiplet in
Mn-$[3\times 3]$ it is $6.6\times10^{-4}$, and for $S=\frac{7}{2}$ it is $29\times10^{-4}$. For comparison,
we list the values of $|B^0_4/D|$ for some of the lowest lying states of several other molecular clusters,
which are considered to show $S$-mixing: $0.2\times10^{-4}$ for Fe$_4$ ($S=5$); $0.5\times10^{-4}$ for
Mn$_{12}$ ($S=10$); $2\times10^{-4}$ for Ni$_4$ ($S=4$); and $5\times10^{-4}$ for the $S=2$ excited state of
the ferric wheel CsFe$_8$ \cite{Liviotti02,Mn12_Fe8_etc,SH_Ni4,OW_NVT}. Apparently, Mn-$[3\times 3]$ shows
the strongest $S$-mixing. In principle, exchange constants may be determined indirectly using Eq.~(1) on the
basis of deviations from the expected $M^2$ behavior. However, this works only for the simplest systems, as
recently demonstrated for a Ni$_4$ SMM \cite{SH_Ni4}. In the present work, the exchange splitting was
determined by direct spectroscopy via EPR.

It is apparent that the effects of $S$-mixing become more important
as the separation between spin multiplets decreases. This argument
can, however, be misleading. For example, the $S = 10$ ground and
first excited $S=9$ multiplets overlap in Mn$_{12}$, yet $S$-mixing
is relatively weak ($|B^0_4/D|=0.5\times10^{-4}$). In contrast,
$S$-mixing is 10 times stronger in Mn-$[3\times 3]$, in spite of the
fact that the $S=\frac{5}{2}$ and $\frac{7}{2}$ states are well
separated in energy (see Fig.~\ref{fig:3}). Evidently, other factors
must be important too, such as the spatial symmetries of the
relevant spin wavefunctions \cite{OW_HighSpin}. Therefore, one
should be careful in judging the importance of $S$-mixing solely on
the basis of the separation of spin levels, or indeed on the ratio
of $J$ and $D$. Understanding this issue is of great importance in
terms of the design of future molecule-based magnets.

%

In summary, we present direct spectroscopic evidence for the mixing of spin wavefunctions in a Mn-$[3\times
3]$ grid through the observation of an otherwise forbidden inter-spin-multiplet EPR transition. Detailed
analysis in terms of a generic spin Hamiltonian provides tremendous insights into the combined effects of
exchange and anisotropy in a prototypical molecular nanomagnet.

%

\begin{acknowledgments}
We acknowledge EC-RTN-QUEMOLNA, contract n$^\circ$
MRTN-CT-2003-504880, the Swiss National Science Foundation, and the
US National Science Foundation (DMR0239481) for financial support.
\end{acknowledgments}

%

%

\begin{references}
\bibitem{dipdip}
Intramolecular dipolar interactions are not negligible, but have
similar effects to the single-ion anisotropy and are captured by the
$\hat{S}^2_{iz}$ terms. Thus, $D$ values should be understood as
including both anisotropy contributions.
\end{references}
\end{document}